\documentclass[letter,scriptaddress,twocolumn,showkeys]{revtex4}

	\usepackage{amsmath}
	\usepackage[sort&compress]{natbib}
	\usepackage{makeidx}
	\usepackage{amsfonts}
	\usepackage[ansinew]{inputenc}
	\usepackage[usenames,dvipsnames]{pstricks}
	\usepackage{subfigure}
	\usepackage{epsfig}
	\usepackage{pst-grad} 
	\usepackage{pst-plot} 
	\usepackage[colorlinks,hyperindex]{hyperref}
	\hypersetup
	{
		colorlinks,%
		citecolor=blue,%
		linkcolor=blue,%
		urlcolor=black,%
	}

\newcommand{\beqn}{\begin{equation}}
\newcommand{\eeqn}{\end{equation}}
\newcommand{\beqna}{\begin{eqnarray}}
\newcommand{\eeqna}{\end{eqnarray}}
\makeindex
\begin{document}
\title{Simulation study of elliptic flow of charged hadrons produced in Au + Au collisions at the Facility for Antiproton and Ion Research}
\author{S. Sarkar, P. Mali and A. Mukhopadhyay}
\email{amitabha$_$62@rediffmail.com}
\affiliation{Department of Physics, University of North Bengal, Siliguri 734013, West Bengal, India}
\begin{abstract}
Centrality and system geometry dependence of azimuthal anisotropy of charged hadrons measured in terms of the elliptic flow parameter are investigated using Au+Au event samples at incident beam energy $20$A GeV and $40$A GeV generated by Ultrarelativistic Quantum Molecular Dynamics and A Multiphase Transport models. The Monte Carlo Glauber model is employed to estimate the eccentricity of the overlapping zone at an early stage of the collisions. Anisotropies present both in the particle multiplicity distribution and in the kinetic radial expansion are examined by using standard statistical and phenomenological methods. In the context of upcoming Compressed Baryonic Matter experiment to be held at the Facility for Antiproton and Ion Research, the present set of simulated results provide us not only with an opportunity to examine the expected collective behavior of hadronic matter at high baryon density and moderate temperature environment, but when compared with similar results obtained from Relativistic Heavy Ion Collider and Large Hadron Collider experiments, they also allow us to investigate how anisotropy of hadronic matter may differ or agree with its low baryon density and high temperature counterpart. 
\end{abstract}
\maketitle
\section{Introduction}
\label{Intro}
It is widely accepted that studies on anisotropic azimuthal distribution of final state particles can be used to explore collective fluid like behavior of hadronic matter produced under extreme thermodynamic conditions in high-energy heavy-ion collisions \cite{Adcox05,Adams05,Adler05,Alver07,Aamodt10}. More precisely, the second harmonic coefficient of an azimuthal Fourier decomposition of particle multiplicity distribution, also known as the elliptic flow parameter $(v_2)$, is of special interest \cite{Olli92,Volo96}. The $v_2$ parameter allows us to critically examine the evolution of the early stage of a high-energy collision between two nuclei \cite{Sorge97}. Large $v_2$ values obtained in experiments using facilities like Relativistic Heavy Ion Collider (RHIC) \cite{Adler05,Alver07, Adam12} and Large Hadron Collider (LHC) \cite{Aamodt10,Abelev15,Adam16}, lead us to conclude that strong anisotropies are present in the azimuthal distribution of particle multiplicities and the strongly interacting partonic and/or hadronic matter produced in collisions between heavy nuclei exhibits collective effects observed only in fluid like states. The dynamics of such a state of matter can be explained by hydrodynamic calculations \cite{Hirano06,Luzum09} as well as by transport models with parton scattering cross-section(s) much larger than that predicted by perturbative Quantum Chromodynamics (pQCD) \cite{Adam12}. However, depending upon the collision energy involved, such models may require some fine tuning \cite{Armesto08}. Most of the observations on collective behavior of QCD matter pertain to high temperature and low net baryon density limits in the QCD phase diagram. On the other hand, such behavior at high baryon density and low or moderate temperature, experimentally is not yet well explored. The upcoming Compressed Baryonic Matter (CBM) experiment \cite{CBM} scheduled to be held sometime around 2018-'19 at the Facility for Anti-proton and Ion Research (FAIR), is expected to provide us with a detailed experimental database on QCD matter at high baryon density and low or moderate temperature. The proposed incident energy range of the CBM heavy-ion programme is $10 - 40$ GeV per nucleon in the laboratory system, which seems to be capable of creating a matter density of about $6 - 12$ times the normal nuclear density at the central particle producing rapidity region \cite{Stoc86}.\\ 
\\
As mentioned above, the flow coefficients are generally measured from the Fourier expansion of the azimuthal distribution of the particle number density
\begin{equation}
\frac{d^3N}{d^3p} = \frac{d^2N}{2\pi E\, p_{_T} dp_{_T} dy} \left[1+\sum_{i=1}^N 2\,v_n \cos\{n(\phi-\psi_{_{RP}})\}\right]
  \label{d3N}
\end{equation}
where $E,~p_{_T},~y,$ and $\phi$ are, respectively the energy, transverse momentum, rapidity and azimuthal angle of the produced particles, while $\psi_{_{RP}}$ is the azimuthal angle of the reaction plane, which is defined as the plane spanned by the impact parameter vector and the incident beam direction \cite{Volo96,Posk98}. If the impact parameter vector is taken along the $x$ direction and the beam along the $z$ direction, then the reaction plane is nothing but the $(p_x,\,p_z)$ plane. Under such circumstances the reaction plane angle $\Psi_{_{RP}}$ becomes zero. Consequently, the $n$th harmonics of the underlying distribution $(v_n)$ reduces to 
\begin{equation}
v_n = \left< \cos(n\phi) \right>
\label{v_n}
\end{equation}
The elliptic flow parameter $v_2$ characterizes the anisotropy in the particle distribution in the momentum space. In Eq.\,(\ref{v_n}) the symbol $\left< ~\right>$ refers to an averaging over all particles considered in the sample, which in the present case are the charged hadrons. For particle number distribution the coefficient $v_2$ reads as 
\begin{equation}
  v_2 = \left< \frac{p_x^2 - p_y^2}{p_x^2 + p_y^2} \right>
  \label{v_2}
\end{equation} 
where $p_x$ and $p_y$ respectively, are the $x$ and $y$-components of the particle momentum. From the RHIC data it has been numerically established that the final stage momentum anisotropy results from initial spatial anisotropy $(\varepsilon)$ of the overlapping volume of two interacting nuclei
\begin{equation}
  \varepsilon = \left< \frac{x^2 - y^2}{x^2 + y^2} \right>
  \label{epsilon}
\end{equation} 
that cannot be directly measured in an experiment \cite{Back}. Here $x,\,y$ are the respective transverse space co-ordinates of participating nucleons in and out of the reaction plane. Fortunately, the elliptic flow parameter $(v_2)$ of charged hadrons at least in mid-central collisions holds a proportionality with the initial spatial eccentricity of the intermediate `fireball' created in collision between two heavy nuclei at high energies. To some extent the linear dependence of $v_2$ on $\varepsilon$ is phenomenological \cite{Kolb00,Bhalerao05}. However, such a linear dependence is found to be consistent with the hydrodynamical calculations with a proper parametrization of the speed of sound in the fireball medium \cite{Bhalerao05,Karp14}.\\ 
\\
One should remember that the initial geometrical anisotropy of the overlapping region in non-central collisions may lead to anisotropy in the kinetic radial expansion as well \cite{Ollitrault98}. The kinetic radial expansion and the asymmetry therein, if any, can be measured from the distribution of total transverse momentum and average transverse momentum \cite{Li2012}. For a large sample of events the total transverse momentum $\left<P_{_T}(\phi_m) \right>$ in the $m$-th azimuthal bin is introduced as
\begin{equation}
	\left< P_{_T}(\phi_m) \right> = \frac{1}{N_{ev}} \sum_{j=1}^{N_{ev}}\,\sum_{i=1}^{n_{_m}} p_{_{T,i}}(\phi_{_m})
	\label{avg-rap}
\end{equation}
where $p_{_{T,i}}(\phi_m)$ is the transverse momentum of the $i$-th particle, $n_{_m}$ is the total number of particles in the $m$-th bin, $N_{ev}$ is the number of events under consideration and $\left<~\right>$ denotes an averaging over events. An azimuthal distribution of $\left< P_{_T}(\phi_m) \right>$ contains information of multiplicity as well as of radial expansion. On the other hand by taking an average over particle number  
\begin{equation}
	\left<\left< p_{_T}(\phi_m) \right> \right> = \frac{1}{N_{ev}}\sum_{j=1}^{N_{ev}} \frac{1}{N_m}\sum_{i=1}^{N_m} p_{_{T,i}}(\phi_m)
	\label{mean-rap}
\end{equation}	
in the mean transverse momentum $\left<\left< p_{_T}(\phi_m) \right> \right>$ the multiplicity influences can be significantly reduced, and the corresponding distribution measures only radial expansion.\\
\\
Our present understanding of collective flow at FAIR energy region is constrained by nonavailability of experimental data. Though the Alternating Gradient Synchrotron (AGS), Super Proton Synchrotron (SPS) and some lowest energy RHIC measurements \cite{Adam12,Agak12,Adam13} provide us with some sort of a database, it is nevertheless necessary to scan a much wider range of collision energies involving different colliding systems that may be used to study baryon rich hadronic matter where high degree of nuclear stopping is expected. In absence of experimental data, simulations that are successful in describing certain phenomenon like collective flow, can provide us with such useful information as to what can be expected in future experiments. Such an exercise will not only help us to understand the dynamics of the system but will also provide important clues that might constrain the models and theories to be used to characterize a baryon rich `fireball'. Keeping this in mind, in this article we have studied some basic aspects of elliptic flow of charged particles produced in Au+Au collisions at the FAIR energies using the Ultra-relativistic Quantum Molecular Dynamics (UrQMD) \cite{UrQMD} and A Multiphase Transport (AMPT) \cite{AMPT, Chen05} model. The Monte Carlo Glauber (MCG) model \cite{Miller07} is employed to characterize the collision geometry at an early stage of the evolution of an $AA$ collision. The latest available version of these models are used to simulate symmetric fixed target nuclear (Au+Au) collisions at incident energies $E_{\rm Lab}=20$A GeV and $40$A GeV. The main motivation behind this kind of simulation based analysis is to examine how different flow parameters are expected to behave in a moderate temperature, baryon rich environment, and in what respects are they similar to or different from a high temperature and almost baryon free `fireball' created in RHIC and/or LHC experiments. The results at high energy density and high temperature in an almost baryon free condition obtained from the RHIC and LHC experiments are available in literature. We shall see that it is worthwhile to compare and supplement the RHIC and LHC results with those obtained from the present analysis of simulated data. Our investigation also covers the impact of system geometry on the differential elliptic flow, and discuss the effects of different averaging schemes of $v_2$ \cite{Li10}. The paper is organized as follows. In Sec.\,\ref{Model} we provide a very brief review of the models used in this work i.e. the UrQMD, AMPT and MCG model. In Sec.\,\ref{Result} we present the results of this analysis and the paper is critically summarized in Sec.\,\ref{Summary}.
\section{Brief description of the models}
\label{Model}
\subsection{UrQMD model}
The Ultra-relativistic Quantum Molecular Dynamics (UrQMD) \cite{UrQMD} model of high-energy nucleus-nucleus $(AA)$ collisions is based on a microscopic transport theory of covariant propagation of hadrons along their classical trajectories combined with stochastic binary scattering, resonance decay and color string fragmentation. In a transport model, an $AA$ collision is assumed to be a superposition of all possible binary nucleon-nucleon $(NN)$ collisions. An $NN$ collision is allowed if the impact parameter $b$ satisfies the criterion $b < \sqrt{\sigma_{\rm tot}}/\pi$, where the total cross-section $\sigma_{\rm tot}$ depends upon the isospin of interacting nucleons and the $NN$ center of mass energy ($\sqrt{s_{NN}}$) involved in the collision. The Fermi gas model is utilized to describe the projectile and the target nuclei. Therefore, the initial momentum of each nucleon is distributed at random between zero and the local Thomas-Fermi momentum. Each nucleon is described by a Gaussian-shaped density function and the wave packet of each nucleus is then taken as a product of single-nucleon Gaussian functions without invoking the Slater determinant necessary for anti-symmetrization of an identical fermion system. At low energies (typically $\sqrt{s_{NN}} \lesssim 5$ GeV) the interaction dynamics is described in terms of the hadronic degrees of freedom, whereas at higher energies ($\sqrt{s_{NN}} \gtrsim 5$ GeV) excitation of color strings and their subsequent fragmentation into hadrons are taken into account. The collision term in UrQMD Hamiltonian includes more than 50 baryon and 45 mesons species, and the model adequately takes care of re-scattering effects.
\subsection{AMPT model}
A Multi-Phase Transport (AMPT) \cite{AMPT, Chen05} model is an example of a hybrid type of transport model. The initial conditions of AMPT are obtained from the Heavy Ion Jet Interaction Generator (HIJING) \cite{HIJING} that uses a Glauber formalism to determine the positions of participating nucleons. AMPT uses Zhang's Parton Cascade (ZPC) formalism \cite{Zhang98} for fixing the scattering properties of partons. Note that the ZPC model includes only parton-parton elastic scattering with an in-medium cross section derived from pQCD with an effective gluon screening mass taken as a parameter. The hadronization process is settled using either the Lund fragmentation or a quark coalescence scheme. Therefore, the users of AMPT have the choice of using either of these two modes of particle production. In the default configuration the energy of the so called excited strings are not used to create partonic states. Like the Lund string fragmentation model, in AMPT default version the energy is released only in the hadronization stage. On the other hand in the string melting configuration, AMPT converts (or melts) all the excited strings into partons (e.g., mesons $\to$ quark and anti-quark pair, baryons $\to$ three quarks etc.). The scattering of the partons are based on parton cascade model \cite{AMPT,Wei02}. At the end of each interaction the leftover quarks/partons are combined into either mesons or baryons through the quark coalescence mechanism.
\subsection{Monte Carlo Glauber model}
The Monte Carlo Glauber (MCG) model is a useful tool to estimate the geometrical configurations of a pair of colliding nuclei. The model operates in two steps, (i) determination of nuclear positions in each nucleus by some stochastic approach and (ii) evaluation of collision properties of the colliding nuclei \cite{Lud86,Shor89}. The position of each nucleon in the nucleus is described by a smooth quantum mechanical single-particle probability density function $\rho$. At least for the closed and near closed shell nuclei such as Au the probability distribution in polar angle as well as in azimuthal angle is taken to be uniform. On the other hand, the radial distribution function is constrained from the nuclear charged density measurement \cite{Vries87}, and is typically characterized by the Fermi distribution,
\begin{equation}
  \rho(r) = \frac{\rho_{_0} }{1+ \exp\left[ (r-R)/a \right]}
\end{equation} 
Here the nuclear radius $R$ and the skin depth $a$ are estimated from low energy electron scattering experiments. The overall normalization parameter $\rho_0$ (nucleon density) is not relevant for this calculation. In order to optimize the nuclear dimension, one may require to set a minimum inter-nucleon separation $(d_{\min})$ between the centers of the nucleons. In this model at relativistic energies the nucleons are assumed to travel along the beam direction throughout the reaction process (eikonal approximation), so that their transverse degrees of freedom are negligible during the time span when the colliding nuclei pass through each other. The impact parameter of each collision is taken at random from a distribution like $dN/db \propto b$ with a large maximum limit $b_{\max}(\simeq 20$ fm. say). In the co-ordinate space $\{x,y,z\}$ the centers of the colliding nuclei are taken at $\{+b/2,0,0\}$ and $\{-b/2,0,0\}$. Due to this conversion the reaction plane is specified by the impact parameter vector and the beam direction, i.e. along the $x$ and $z$-axes, while the $\{x,y\}$-plane denotes the transverse plane, and $\phi=\tan^{-1}\left(p_y/p_x\right)$ the azimuthal angle. An $AA$ interaction in the MCG model is fully specified by the inelastic $NN$ cross-section ($\sigma_{\rm NN}$) that depends only on the energy of the collision. In the MCG model an $NN$ interaction takes place if the Euclidean transverse distance $(D)$ between the centers of any pair of nucleons is less than $D=\sqrt{\sigma_{\rm NN}/\pi}$.\\
\\
For our analysis we chose a configuration of the MCG model which is similar to that used in the PHOBOS experiment \cite{Back}. For Au+Au collision the parameters are selected as, nuclear radius $R=6.38$ fm, skin depth $a=0.535$ fm, $d_{\min}=0$, and the $NN$ cross-section $\sigma_{NN}=30.5$ mb at $20$A GeV and $30.8$ mb at $40$A GeV. With these specifications we compute the number of participating nucleons $N_{\rm part}$, and the total number of $NN$ collisions $(N_{\rm coll})$ on an event by event basis taking the collision impact parameter value from the UrQMD and AMPT models. Note that $N_{\rm part}$ counts the number of nucleons which struck at least once during the collision process, while $N_{\rm coll}$ counts only the inter-nucleon collisions if the colliding nucleons do not stem from the same nucleus. The MCG model records the status and position of each nucleon in the event which are used latter to calculate the spatial eccentricity $(\varepsilon)$ of the collision, and to identify the centrality of a collision in terms of the number of participating nucleons $N_{\rm part}$. The MCG model assumes that the minor axis of the overlap ellipsoid is directed along the impact parameter vector, which as mentioned above is considered as the $x$ direction, while the $y$ direction lies perpendicular to the impact parameter in the transverse plane. Thus the eccentricity is defined as \cite{Sorg99}
\begin{equation}
  \varepsilon_{\rm std} = \frac{\sigma_y^2 - \sigma_x^2}{\sigma_y^2 + \sigma_x^2}
\end{equation} 
where $\sigma_x^2=\left<x^2\right>-\left<x\right>^2$ and $\sigma_y^2=\left<y^2\right>-\left<y\right>^2$ are the variances of the nucleon distribution along the $x$ and $y$ directions in a given interaction. The spatial eccentricity defined in this way is called the `standard' or `reaction plane' eccentricity. There is another choice of $\varepsilon$ that arises because of the mismatch between the minor axis of the ellipsoid in the transverse plane created by the participating nucleons and the geometrical impact parameter vector. This choice of $\varepsilon$ puts more emphasis on the number of participating nucleons in an interaction, and it is termed as the participant eccentricity $\varepsilon_{\rm part}$. The simplest and efficient way to calculate $\varepsilon_{\rm part}$ is to perform a principal axis transformation \,--\, rotate the coordinate system to make $\sigma_x$ minimum. Then in terms of the original coordinate system $\varepsilon_{\rm part}$ is given by \cite{alver08}
\begin{equation}
  \varepsilon_{\rm part} = \frac{\sqrt{(\sigma_y^2 - \sigma_x^2)^2 + 4(\sigma_{xy})^2}}{\sigma_y^2 + \sigma_x^2}
\end{equation}    
where $\sigma_{xy} = \left< xy \right> - \left<x\right> \left<y\right>$. For very heavy interacting nuclei such as Au+Au system, the average values of $\varepsilon_{\rm std}$ and $\varepsilon_{\rm part}$, but for the most peripheral collisions are quite similar. However, for lighter systems like Cu+Cu, significant differences are observed between the two measures of $\varepsilon$ at all centralities \cite{Alver07}.    
\section{Results and discussion}
\begin{figure}[tbh]
\centering
\vspace{-1.2cm}
\includegraphics[height=8cm, width=9cm]{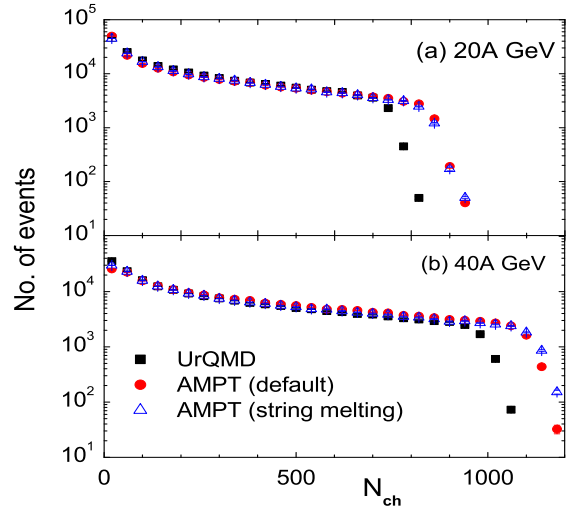}
\vspace{-.8cm}
\caption{(Color online) Charged hadron multiplicity distribution in Au+Au collisions at $E_{\rm Lab}=20$A and $40$A GeV.}
\label{multi}
\end{figure}

\begin{figure}[tbh]
\centering
\vspace{-.8cm}
\includegraphics[width=0.5\textwidth]{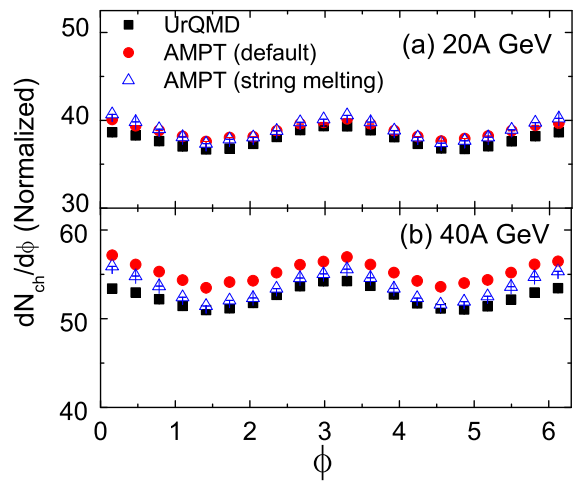}
\vspace{-.8cm}
\caption{(Color online) Charged hadron distribution in the azimuthal plane in Au+Au collisions at $E_{\rm Lab}=20$A and $40$A GeV.}
\label{phi_dist}
\end{figure}

\begin{figure}[tbh]
\centering
\vspace{-.8cm}
\includegraphics[width=0.5\textwidth]{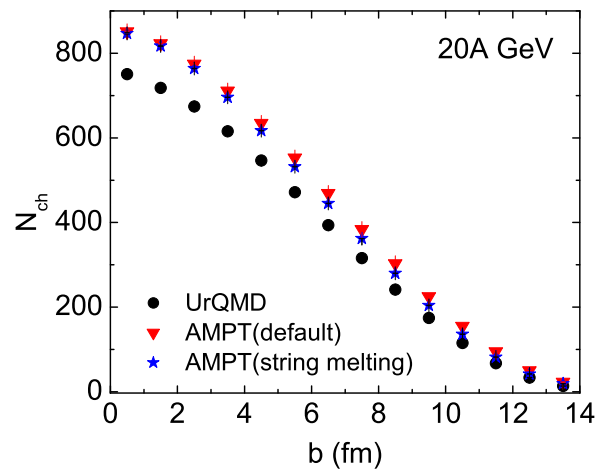}
\vspace{-.7cm}
\caption{(Color online) Charged hadron multiplicity $N_{\rm ch}$ as a function of impact parameter $b$ in Au+Au collisions at $E_{\rm Lab}=20$A GeV. Note that at higher centrality the multiplicity in UrQMD is lower than that in AMPT.}
\label{nch-b}
\end{figure}

\begin{figure}[tbh]
\centering
\vspace{-.8cm}
\includegraphics[width=0.5\textwidth]{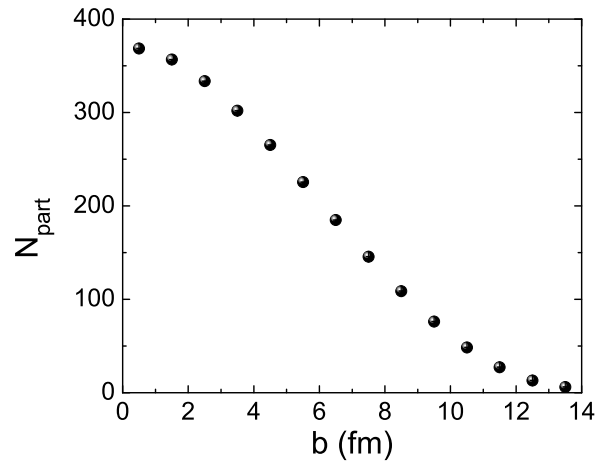}
\vspace{-.7cm}
\caption{Impact parameter dependence of the number of participating nucleons $N_{\rm part}$ in Au+Au collisions in the Monte Carlo Glauber model.}
\label{npart-b}
\end{figure}

We start by plotting the charged hadron multiplicity $(N_{ch})$ distributions  [Fig.\,\ref{multi}] in Au+Au simulated events at $E_{\rm Lab}=20$A GeV and $40$A GeV. Each sample consists of $0.2$ million events. For all three models used in this investigation the nature of multiplicity distributions are more or less similar. However, the maximum multiplicity in UrQMD is less than that obtained from AMPT. In Fig.\,\ref{phi_dist} we plot the azimuthal angle distributions of all charged hadrons produced in the Au+Au collisions considered in the present investigation. We observe the presence of anisotropy in each such plot, and also find that the particle density $\left(N_{ev}^{-1}\,dN_{ch}/d\phi\right)$ i.e., the number of particles per unit $\phi$ interval, is consistently less in UrQMD generated event sample than in AMPT. The differences are more in the $40$A GeV event sample.
In Fig.\,\ref{nch-b} we plot the charged hadron multiplicity $N_{\rm ch}$ as a function of impact parameter $b$ in Au+Au collisions at $E_{\rm Lab}=20$A GeV for all the model simulated samples used in this investigation. At a given impact parameter we see that compared to the AMPT model the UrQMD model produces less number of charged hadrons, the difference however deceases with increasing impact parameter. At highest centrality the multiplicity difference between the two models counts to be as large as $100$. At the present scale of $E_{\rm Lab}$ the nature of $N_{\rm ch}$ versus $b$ plot is found to be almost energy independent. One can either consider the impact parameter (applicable only for model simulation) or the event multiplicity as a measure of the collision centrality. But the results in Fig.\,\ref{nch-b} indicates that $N_{\rm ch}$ as a function of $b$ is not identical for both the models under consideration. Further, whenever a centrality dependence of some result from two different models are to be compared, a common algorithm for the centrality measurement should be used. We use the Monte Carlo Glauber (MCG) model for this purpose. For a given impact parameter, which is obtained from the event generator, we find the total number of participating nucleons $N_{\rm part}$ in the interaction and their coordinate records from the MCG model. $N_{\rm part}$ quantifies the centrality of the collision and the coordinates of all $N_{\rm part}$ nucleons are used to evaluate the eccentricity of the ellipsoid shaped reaction zone. For the UrQMD generated Au+Au collisions at $20$A GeV the $N_{\rm part}$ versus $b$ plot is shown in Fig.\,\ref{npart-b}. We do not find any visible difference between the UrQMD and the corresponding AMPT model generated values and hence the AMPT plot is not shown here. In this regard the $E_{\rm Lab}=40$A GeV results for different models are not significantly different either. From Fig.\,\ref{nch-b} and Fig.\,\ref{npart-b} it is obvious that the correlation between $N_{\rm ch}$ and $N_{\rm part}$ is almost linear and hence the parameter $N_{\rm part}$ can be a proper choice to determine collision centrality.
\label{Result}
\subsection{Centrality dependence of $v_{_2}$}
\begin{figure}[tbh]
\centering
\vspace{-.4cm}
\includegraphics[width=0.51\textwidth]{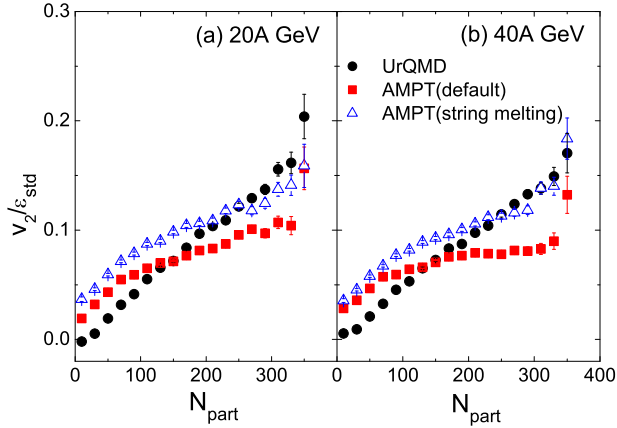}
\vspace{-1cm}
\caption{(Color online) $v_2/\varepsilon_{\rm std}$ as a function of $N_{\rm part}$ for the charged hadrons in Au+Au collisions at (a) $E_{\rm Lab} = 20$A GeV and (b) $E_{\rm Lab} = 40$A GeV. Statistical errors are shown.}
\label{v2estd}
\end{figure}

\begin{figure}[tbh]
\centering
\vspace{-.5cm}
\includegraphics[width=0.51\textwidth]{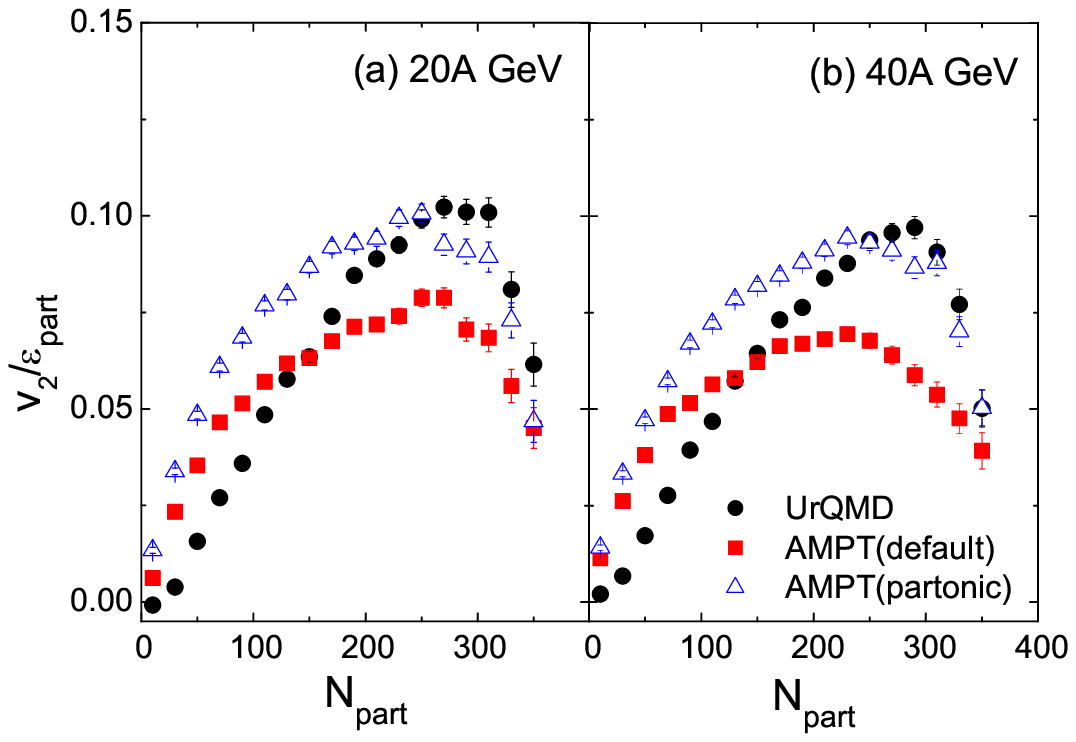}
\vspace{-1cm}
\caption{(Color online) $v_2/\varepsilon_{part}$ as a function of $N_{\rm part}$ for the charged hadrons in Au+Au collisions at (a) $E_{\rm Lab} = 20$A GeV and (b) $E_{\rm Lab} = 40$A GeV. Statistical errors are shown.}
\label{v2epart}
\end{figure}

The centrality dependence of elliptic flow can provide us with valuable information on the nature of rescattering (whether hadronic or partonic) and the degree of thermalization achieved in the `fireball' created in an $AA$ collision \cite{Volo00}. For extreme peripheral and central collisions the values of elliptic flow parameter are usually smaller than those observed in mid-central collisions. This is also observed in all calculations where transport models are used \cite{Zhu05,Lu06}, and also in low energy collisions especially at AGS energies \cite{Barr97}. The observation could be explained in terms of the geometry and pressure gradient produced at the early stage of interactions. Also the observation is consistent with the low density limit of the hydrodynamical model. In addition, the effect of shadowing by spectator nucleons plays a crucial role for suppressing the elliptic flow in peripheral collisions. As we move from AGS to SPS \cite{Barr95}, then to RHIC \cite{Adam12} and finally to the LHC experiments \cite{Abelev15, Adam16}, the flow peak shifts toward more peripheral collisions. With increasing collision energy the physics of central collisions evolves from hadronic to partonic degrees of freedom, resulting thereby a higher multiplicity and correspondingly a weaker flow. At SPS the maximum measured value of elliptic flow ($v_2 \leq 0.04$) is significantly less than the hydrodynamic prediction ($v_2 \sim 0.1$) \cite{Olli92,Kolb99}, and the observed centrality dependence of $v_2$ also does not require a hydrodynamical explanation. On the other hand the RHIC (top energy) data, especially for the mid-central collisions up to $p_{_T} \approx 1.5$ GeV/c, can be described by a hydrodynamic calculation \cite{Kolb01}. \\
\\
\begin{figure}[tbh]
\centering
\vspace{-.5cm}
\includegraphics[width=0.5\textwidth]{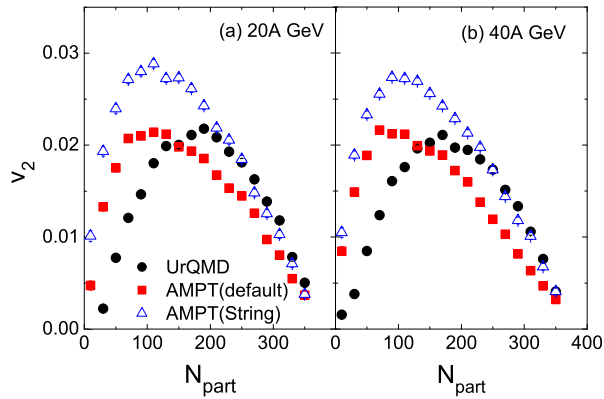}
\vspace{-.7cm}
\caption{(Color online) Centrality dependence of elliptic flow of charged hadrons in Au+Au collision at (a) $E_{\rm Lab} = 20$A GeV and (b) $E_{\rm Lab}=40$A GeV.}
\label{v2-npart}
\end{figure}
\begin{figure}[tbh]
\centering
\vspace{-.5cm}
\includegraphics[width=0.5\textwidth]{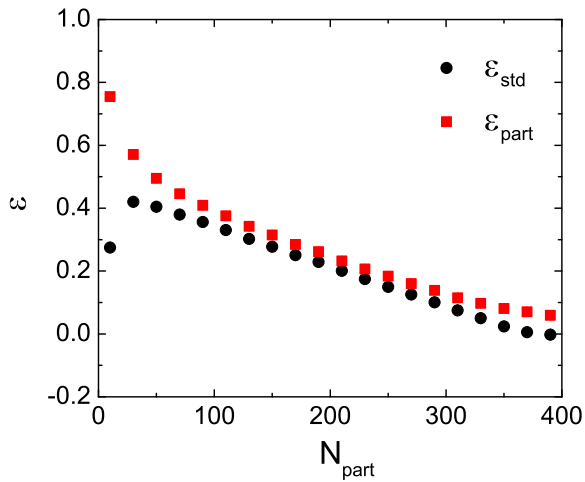}
\vspace{-.7cm}
\caption{(Color online) Centrality dependence of the average eccentricity defined in two different ways, $\varepsilon_{\rm std}$ and $\varepsilon_{\rm part}$, for Au+Au events simulated by the MCG model.}
\label{e-npart}
\end{figure}
We compute the elliptic flow parameter $v_2$ as a function of collision centrality $N_{\rm part}$ for all event samples used in this analysis. The results are presented in Fig.\,\ref{v2-npart}(a) for $E_{\rm Lab}=20$A GeV and in Fig.\,\ref{v2-npart}(b) for $E_{\rm Lab}=40$A GeV collisions. It is seen that the centrality dependence of $v_2$ calculated from any of the models hardly changes as the beam energy is increased from $20$ to $40$ GeV per nucleon. The elliptic flow parameter becomes vanishingly small in most central and most peripheral collisions. Using the UrQMD and the AMPT models a similar kind of observation was also made in Au+Au collisions at $\sqrt{s_{NN}}=200$ GeV \cite{Zhu05,Zhou10}. The UrQMD and AMPT (default) generated $v_2$ values are more or less same, while the AMPT (string melting) model produces significantly higher values at both energies. Another interesting observation of Fig.\,\ref{v2-npart} is that, the models yield noticeably different values of $v_2$ in the mid-central and peripheral events ($N_{\rm part} \leq 200$), whereas in more central region $(N_{\rm part} > 200)$ corresponding $v_2$ values are not much different. The maximum of $v_2$ that is obtained for the UrQMD model at $N_{\rm part} \approx 200$ corresponds to $b \approx 5.5-6.5$ fm, which is of the same order of the colliding Au nuclei. At same $E_{\rm Lab}$ in the $v_2$ vs. $N_{\rm part}$ plot, the peak values for the AMPT generated samples are observed at comparatively lower values of $N_{\rm part}$. All these observations indicate that in central collisions the shape of the overlapping system largely dominates the flow characteristics, while in mid-central collisions hadronic and partonic (re)scattering might have taken a leading role. As discussed in the previous section, the fluctuation in the number of participating nucleons and their positions may have a formidable impact on the spatial eccentricity calculated by the MCG model. So before we study the effects of eccentricity on the flow results, it is important to note the centrality dependence of the eccentricity itself. The MCG model simulated results on such an analysis for Au+Au collisions at $E_{\rm lab}=20$A GeV is presented in Fig.\,\ref{e-npart} and for some selected centrality classes the average value of $N_{\rm part}$, $\varepsilon_{\rm std}$ and $\varepsilon_{\rm part}$ are listed in Table \ref{table}. The table as well as the figure show that in Au+Au system the participant eccentricity $\varepsilon_{\rm part}$ always underestimates the standard geometrical eccentricity $\varepsilon_{\rm std}$. The difference between these two measures of $\varepsilon$ is $\sim 20$\% in the $N_{\rm part} \approx 100-300$ region. The differences are magnified in extreme central and peripheral collisions. In most central collisions we find $\varepsilon_{\rm std}\approx 0$, as the definition demands, but $\varepsilon_{\rm part} > 0$. For $N_{\rm part}$ below $30$ we observe a small dip in the variation of $\varepsilon_{\rm std}$ that can be attributed to a very small number of participating nucleons, sometimes even as small as one each from the colliding nuclei, lying in the reaction plane of extreme peripheral collisions. Our overall observation on $N_{\rm part}$ dependence of $\varepsilon$ matches with that observed in Au+Au collisions at $\sqrt{s_{NN}} = 200$ GeV \cite{Alver07}. It should be noted that the MCG model operates with several experimentally determined parameters. Therefore, the model calculated values of eccentricity are not free from systematic errors incurred by the input parameters. Alver {\it et al.} \cite{Alver07} attempted to estimate the maximum possible systematic error in $\varepsilon$. According to the MCG model the eccentricity values are of 90\% confidence level.\\ 
\begin{table}
\vspace{-.2cm}
\caption{Mean participant nucleon number $\left< N_{\rm part} \right>$, average values of $\varepsilon_{\rm std}$ and $\varepsilon_{\rm part}$ at different centrality bins in Au+Au collisions at $E_{\rm Lab}=20$A GeV and 40A GeV, obtained from the MCG model.}
\centering
\begin{tabular}{lllll}
\hline\hline
Centrality & $~~~\langle N_{\rm part} \rangle$~~~ & $~~~\langle\varepsilon_{\rm std}\rangle$~~~~~~~ & $~~~~\langle \varepsilon_{\rm part} \rangle$ \\ [0.5ex]
\hline
Au+Au at &20A GeV \\
 ~~0--10\%&  317.5$\pm$0.2 & 0.068$\pm$0.001 & 0.113$\pm$0.001 \\ 
 10--20\% &  230.8$\pm$0.2 & 0.175$\pm$0.001 & 0.208$\pm$0.001 \\ 
 20--30\% &  166.5$\pm$0.1 & 0.259$\pm$0.001 & 0.293$\pm$0.001 \\ 
 30--40\% &  118.0$\pm$0.1 & 0.322$\pm$0.001 & 0.365$\pm$0.001 \\ 
 40--50\% &  80.4$\pm$0.1 & 0.373$\pm$0.001 & 0.429$\pm$0.002 \\ 
 50--60\% &  52.4$\pm$0.1 & 0.409$\pm$0.002 & 0.492$\pm$0.001 \\ 
 60--70\% &  31.9$\pm$0.1 & 0.430$\pm$0.002 & 0.564$\pm$0.002 \\ 
 70--80\% &  17.8$\pm$0.1 & 0.412$\pm$0.003 & 0.645$\pm$0.002 \\ 
Au+Au at& 40A GeV \\
  0-10\%  & 318.3$\pm$0.2 & 0.065$\pm$0.001 & 0.111$\pm$ 0.001\\ 
 10-20\% & 231.1$\pm$0.2 & 0.177$\pm$0.001 & 0.209$\pm$ 0.001\\
 20-30\% & 166.9$\pm$0.1 & 0.258$\pm$0.001 & 0.292$\pm$ 0.001\\ 
 30-40\% & 118.4$\pm$0.1 & 0.323$\pm$0.001 & 0.366$\pm$ 0.001\\ 
 40-50\% &  80.8$\pm$0.1 & 0.370$\pm$0.001 & 0.427$\pm$ 0.001\\ 
 50-60\% &  52.6$\pm$0.1 & 0.408$\pm$0.001 & 0.493$\pm$ 0.001\\ 
 60-70\% &  32.2$\pm$0.1 & 0.425$\pm$0.002 & 0.564$\pm$ 0.001\\ 
 70-80\% &  18.3$\pm$0.1 & 0.408$\pm$0.002 & 0.642$\pm$ 0.001\\ 
\hline\hline
\end{tabular}
\label{table}
\end{table}      
\\
We now study the impact of the shape and/or size of the system created at the early stage of collision on the elliptic flow coefficient $v_2$. This is done by normalizing $v_2$ obtained for a given centrality range by the eccentricity for the same centrality range. Figure\,\ref{v2estd} shows the eccentricity-scaled elliptic flow $v_2/\varepsilon_{\rm std}$ plotted against $N_{\rm part}$ for all the three model simulated Au+Au events at $E_{\rm Lab}=20$A GeV and $40$A GeV. Notice that the bell shaped pattern of $v_2$ seen in Fig.\,\ref{v2-npart} disappears when $v_2$ is divided by $\varepsilon_{\rm std}$. In Fig.\,\ref{v2estd} the $v_2/\varepsilon_{\rm std}$ values for UrQMD and AMPT (string melting) models are found to increase monotonically with $N_{\rm part}$ over the entire centrality range, and at large $N_{\rm part}( > 200)$ values both models produce nearly same values of $v_2/\varepsilon_{\rm std}$. For the default AMPT model the $v_2/\varepsilon_{\rm std}$ values tend to saturate at $N_{\rm part}>200$. In the peripheral region the UrQMD calculated values are lower than the AMPT calculated values. It is quite surprising that the bell shaped pattern of $v_2$ versus $N_{\rm part}$ plot is now completely washed out and for central collisions we obtain highest value of elliptic flow when it is scaled by eccentricity. We note that, starting from the initial geometry the evolution of $v_2$ is some kind of a quenching process. Therefore, depending on the evolution, it can drastically change after being normalized by eccentricity. Our observation has a similarity with the RHIC Au+Au and Cu+Cu data too \cite{Alver07}. From the above discussion we guess that $\varepsilon_{\rm std}$ may not be an appropriate choice that can effectively represent collision eccentricity. Therefore, we divide $v_2$ by the participant eccentricity $\varepsilon_{\rm part}$ and plot the same in Fig.\,\ref{v2epart}. We find that for all the models used as well as in both the energies under consideration, the highest elliptic flow occurs at around $N_{\rm part} = 250$, i.e. in semi-central collisions. The bell shape of $v_2$ versus $N_{\rm part}$ plots is partially retrieved. However, the peak positions are now shifted toward slightly higher $N_{\rm part}$ value, while the skewness of the plots get inverted. To some extent our observations are in contrast with the RHIC results at $\sqrt{s_{NN}}= 62.4$ and $200$ GeV \cite{Alver07,Manly06}, where the ratio $v_2/\varepsilon_{\rm part}$ was found to saturate for $N_{\rm part}$ above $100$. It should be noted that ideal hydrodynamical model calculation with hadron cascade can roughly interpret the RHIC data, whereas without hadron cascade the hydrodynamic prediction is more or less similar to the behavior observed in the present case \cite{Aich10}. \\
\begin{figure}[tbh]
\centering
\vspace{-.7cm}
\includegraphics[width=0.5\textwidth]{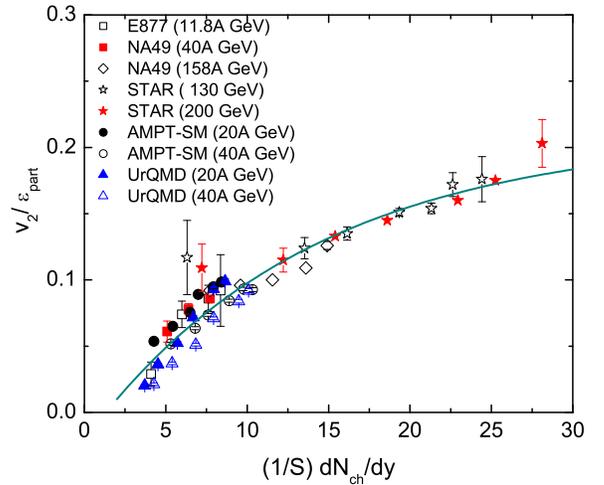}
\vspace{-1cm}
\caption{(Color online) Variation of scaled elliptic flow with scaled particle density. The solid curve is to guide the eye.}
\label{v2e}
\end{figure}
\\
Another important observation of hydrodynamics is that if the equation of state of the model corresponds to a constant sound speed, i.e., there is no phase transition, then the ratio $v_2/\varepsilon$ will be independent of collision centrality \cite{Olli92}. But with a phase transition where sound speed is not constant, hydrodynamical calculation predicts a change in $v_2/\varepsilon$ ratio with centrality. However, such changes are not expected to exceed $10\%$ of the values \cite{Kolb00}. The results of E877 (AGS) \cite{Barr95}, NA49 (SPS) \cite{Posk99,Alt03} and STAR (RHIC) \cite{Adler02} experiments show a smooth linear rising trend of $v_2/\varepsilon$ with increasing centrality. The results are consistent with the so called low density limit (LDL) of hydrodynamical calculation \cite{Volo00,Heis99}, where the mean free path of interaction is at least comparable to the dimension of the nuclear overlap region. This is to some extent identical to the interaction scheme implemented into a transport model where the mean number of hard binary collisions per participant is typically small.
We calculate the ratio $v_2/\varepsilon_{\rm part}$ as a function of charge particle rapidity density $dN_{\rm ch}/dy$ normalized by the transverse area of the overlap region $S = \pi \sqrt{\left<x^2\right> \left< y^2\right>}$ over a wide range of collision centrality $(0-60\%)$ using equal width $(5\%~\mbox{each})$ of centrality interval. This exercise is done for four event samples, two generated by UrQMD and two by AMPT (string melting) models. Our results are graphically presented in Fig.\,\ref{v2e} along with similar results obtained from AGS, SPS and RHIC experiments \cite{Barr95,Posk99,Alt03, Adler02}. To maintain clarity of the diagram the AMPT (default) results and all points corresponding to our calculated results are not plotted in this graph. An approximately proportional relationship like 
\begin{equation}
\frac{v_2}{\varepsilon_{\rm part}}\, \propto \, \frac{1}{S}\,\frac{dN_{\rm ch}}{dy} 
\end{equation}
is reproduced in the cases studied, which suggests that if the effect of nuclear geometry is taken into account, the elliptic flow increases with particle density, at least for a limited centrality interval $(10-60\% ~ \mbox{in our case})$. We must also mention that for a few highest centrality intervals ($0-5\%$ and $5-10\%$) our $v_2/\varepsilon_{\rm part}$ values significantly drop down from the average rising trend with increasing $S^{-1}\,dN_{\rm ch}/dy$ of the rest. The cause of this deviation may be attributed to an altogether different multiparticle production mechanism and/or interation between the particles present in the intermediate `fireball'. The main point is that, as we plot the elliptic flow parameter scaled by the eccentricity against the particle density scaled by transverse overlapping area of the colliding nuclei, at least within a limited centrality interval, which in our case is $10-60\%$, the variation depends only marginally on the collision energy and/or the colliding system involved, thereby indicating some kind of universal nature of the data. Ultimately, this increasing trend should saturate at the hydro limit corresponding to complete thermalization. If the intermediate `fireball' is comparatively dilute, the asymmetry in azimuthal distribution is directly proportional to the number of rescatterings, or equivalently to the particle density in the transverse plane \cite{Posk99,Heis99}. A deviation from the approximately linear scaling observed above over a widely varying collision energies, in real experiments as well as in simulations, will require a different physics of rescattering. Significantly the transport model calculated values presented here are not very much different from the E877 and NA49 results either, which are comparable to the FAIR-CBM energy scale. As mentioned, a transport model prediction should in principle match with the LDL of hydrodynamical calculation. In that sense the results obtained here are not surprising. However, we argue that the results may provide useful information to constrain the equation of state of a highly dense baryonic matter expected to be created at FAIR-CBM experiments. One should keep in mind that there are several sources of uncertainties in the MCG model that might influence the measurement of $S$. The results presented here may contain a significant amount of systematic errors. Also some of the experimental $v_2$ values shown in the plot are calculated under different algorithms e.g., the 4th order cumulant method is used in the STAR and SPS measurement, and the STAR data possess a $p_{_T}$ cut like $0.075 < p_{_T}<2.0$ GeV/c. 
\begin{figure}[t]
\vspace{-0.8cm}
\centering
\includegraphics[width=0.5\textwidth]{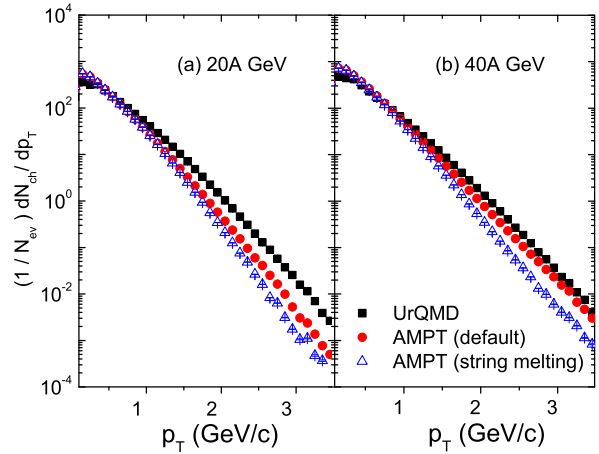}
\vspace{-1cm}
\caption{(Color online) $p_{_T}$ distribution of charged hadrons in Au+Au collisions.}
\label{ptdist}
\vspace{-0.2cm}
\end{figure}
\begin{figure}[tbh]
\vspace{-0.8cm}
\centering
\includegraphics[width=0.5\textwidth]{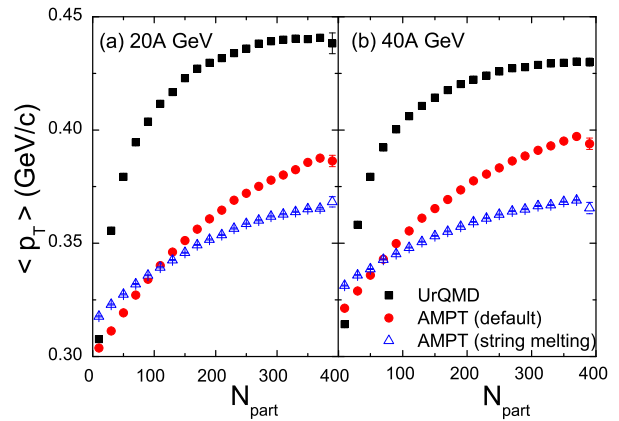}
\vspace{-0.9cm}
\caption{(Color online) Centrality dependence of average transverse momentum.}
\label{avgpt}
\vspace{-0.2cm}
\end{figure}
\subsection{$p_{_T}$ dependence of $v_{_2}$}
\begin{figure*}[tbh]
\vspace{-1cm}
\centering
\includegraphics[width=0.8\textwidth]{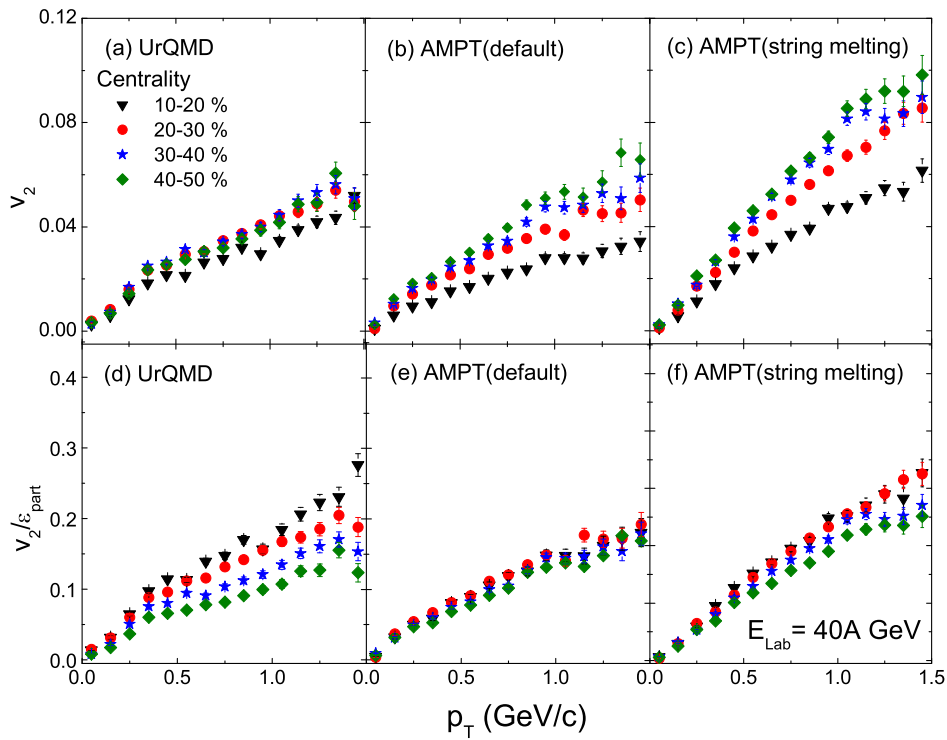}
\vspace{-1cm}
\caption{(Color online) The same as in Fig.\,\ref{v2pt40} but at $E_{\rm Lab}=40$A GeV.}
\label{v2pt40}
\end{figure*}

We have obtained the $p_{_T}$ distributions of charged hadrons produced in Au+Au collisions at $E_{\rm Lab}=20$A GeV and $40$A GeV, and graphically plot them in Fig.\,\ref{ptdist}. As expected we observe an approximately exponential fall in the particle number density with increasing $p_{_T}$. At high $p_{_T}$ region the UrQMD yield is highest and the corresponding slope value of the exponential fall is lowest at both energies. Without any definite indication of thermalization, it would not be proper to extract temperature values from these slopes. We however note that perhaps due to the quark coalescence mechanism of hadronization embedded in AMPT (string melting) model, the $p_{_T}$ distribution is stiffest in the AMPT generated distributions, a feature that should be reflected in large transverse flow of particles. In Fig.\,\ref{avgpt} we have graphically shown a comparison of average $<p_{_T}>$ distribution with collision centrality $N_{\rm part}$ for different models and at the two collision energies considered. At low centrality $<p_{_T}>$ rises with $N_{\rm part}$ which saturates at high centrality. It is interesting also to note that the saturation $<p_{_T}$ values hardly differ at the two collision energies involved, but they are significantly different for the models used -- highest in UrQMD and lowest in AMPT (string melting). Perhaps the quark coalescence mechanism is again responsible for the lowest saturation $<p_{_T}>$ values seen in the AMPT (string melting) model. 
\begin{figure*}[tbh]
\vspace{-1cm}
\centering
\includegraphics[width=0.8\textwidth]{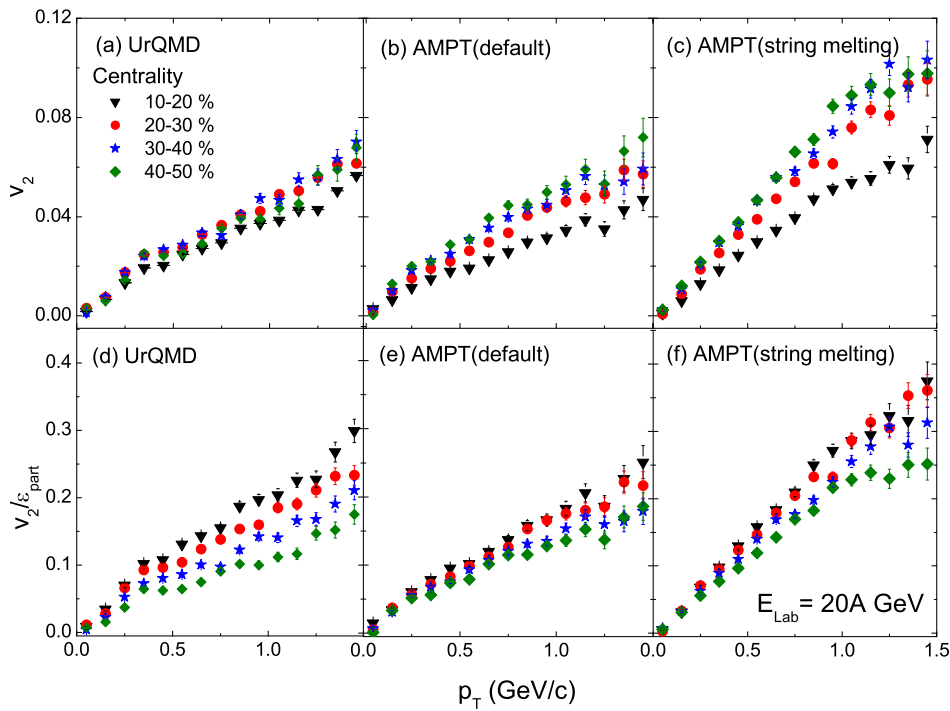}
\vspace{-1cm}
\caption{(Color online) $v_2$ and $v_2/\varepsilon_{\rm part}$ plotted against $p_{_T}$ for charged hadrons in Au+Au collisions at $E_{\rm Lab} = 20$A GeV. For an easy comparison UrQMD and AMPT results are plotted at same scale.}
\label{v2pt20}
\end{figure*}
In the top panels of Fig.\,\ref{v2pt20} and Fig.\,\ref{v2pt40} each, we now plot the elliptic flow parameter against $p_{_T}$ of charged hadrons in Au+Au collisions at $E_{\rm Lab}=20$A GeV and $40$A GeV, respectively. In these figures we plot $v_2$ only for some selected centrality bins, namely $10-20\%$, $20-30\%$, $30-40\%$ and $40-50\%$, since the $v_2$ values for the extreme central and peripheral collisions are significantly dominated by statistical uncertainties. The mean number of participant nucleons $\left< N_{\rm part} \right>$ in these selected centrality bins are given in Table \ref{table}. For an easy reference we plot the results of UrQMD, AMPT (default) and AMPT (string melting) models at the same scale. In all the cases we see that for a given centrality, $v_2$ increases almost linearly with increasing $p_{_T}$, the rate of increase being highest in the AMPT string melting version. The $v_2$ at highest $p_{_T}$ acquires quite high values (more than $0.1$) at least for the AMPT string melting model, but no siginificant difference between $20$A GeV and $40$A GeV results are observed. Also in the studied cases $v_2$ values show a centrality ordering i.e., for a given $p_{_T}$, $v_2$ increases with increasing centrality. However, for $p_{_T}> 1.25$ GeV/c this ordering starts to be affected by the statistical fluctuations. For UrQMD at both energies the ordering is limited only to the first two centrality intervals, while for AMPT it is present in all four. The trends of variation agree well with the same in the low $p_{_T}$ regions of RHIC data \cite{Adams05,Adare10}. Since high $p_{_T}$ hadron production is very much collision energy dependent, and at FAIR energies the high $p_{_T}$ hadrons will be rare, the flow results at large $p_{_T}\,(> 1.5\;\mbox{GeV})$ would not be statistically very significant in our case. In addition the UrQMD and AMPT generated Au+Au events at FAIR energies have been found to follow a mass ordering for $p_{_T} \leq 1$ GeV/c \cite{Partha10} i.e., for a given $p_{_T}$ heavier particle species produces less $v_2$ and vice versa, an aspect that is well reproduced by the hydrodynamical calculation \cite{Kolb01}.
The most direct evidence that $v_2$ is related to spatial asymmetry originating in the early phase of $AA$ interaction is that, $v_2$ at low $p_{_T}$ approximately scales with the initial eccentricity of the collision zone. For reasons discussed above, we use $\varepsilon_{\rm part}$ to scale $v_2$. The model calculated $v_2$ values scaled by $\varepsilon_{\rm part}$ are plotted against $p_{_T}$ for the same centrality intervals as $v_2$ plots in the bottom panels of Fig.\,\ref{v2pt20} and Fig.\,\ref{v2pt40}, respectively for 20A GeV and 40A GeV event samples. The gross characteristics of the $v_2/\varepsilon_{\rm part}$ and $v_2$ plots are more or less similar. At low $p_{_T}$ the eccentricity scaled elliptic flow values are roughly centrality independent but with increasing $p_{_T}$ this centrality independence is broken in all cases but for the AMPT default mode at 40A GeV. We also observe that the centrality ordering of the $v_2$ and $v_2/\varepsilon_{\rm part}$ plots are inverted. The feature is more prominent for the UrQMD generated event samples at both energies. This observation complements previously shown difference between Fig.\,\ref{v2-npart} and Fig.\,\ref{v2epart}. It may be noted that in the RHIC beam energy scan program a similar kind of centrality ordering of the eccentricity scaled $v_2$ was observed \cite{Adam12}. 
\subsection{Average $v_2$}
\begin{figure}[t]
\vspace{-.2cm}
\centering
\includegraphics[width=0.51\textwidth]{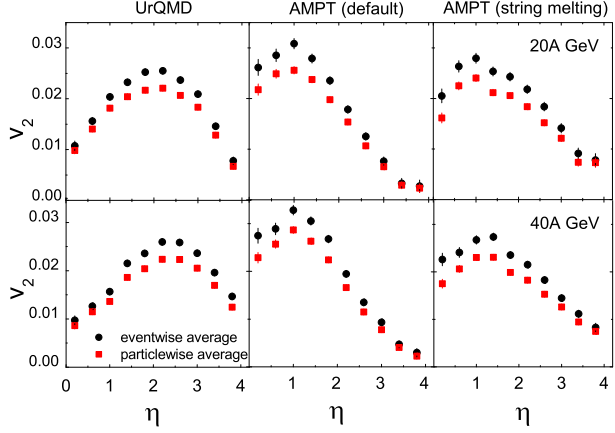}
\vspace{-1cm}
\caption{(Color online) Comparison between particle-wise and event-wise averaged values of $v_2$ plotted against $\eta$ in Au+Au collisions.}
\label{v2-eta}
\end{figure}

The averaging scheme used in Eq.\,\eqref{v_n} and Eq.\eqref{v_2} refers to the so called particle-wise average. As introduced in \cite{Li10} there exists another kind of averaging of $v_n$, which is known as event-wise averaging. Though the particle-wise average is widely accepted in data analysis and theoretical calculations \cite{Chen05}, it does not take the influence of multiplicity fluctuation of the event sample into account. In connection with the elliptic flow measurement the correlation between these two types of averaging schemes is studied in \cite{Li10}. According to \cite{Li10} the event-wise averaged the second harmonic parameter $v_2^e$ reads as
\begin{equation}
  v_2^e = \left< \overline{\cos(2\phi)}\right>_{ev}        
\end{equation}
Here $\overline{\cos(2\phi)}$ represents the average of $\cos(2\phi)$ over all particles in an event, and $\left<~\right>_{ev}$ represents an average over all events present in a sample. Under the same notation the particle-wise average denoted by $v_2^p$ is written as
\begin{equation}
  v_2^p = \left< \overline{\cos(2\phi)}\, N_{ch}\right>_{ev} /\left< N_{ch} \right>_{ev}
\end{equation} 
Note that if $\overline{\cos(2\phi)}$ is independent of multiplicity $N_{ch}$, $v_2^p$ turns into $v_2^e$. This is the case when the harmonic parameter is studied within a very narrow centrality bin. We quantify the difference between particle-wise and event-wise averaged values of the elliptic flow parameter. Fig.\,\ref{v2-eta} displays a sample plot of such analysis, where both the particle-wise and event-wise averaged values of $v_2$ are plotted against pseudorapidity $(\eta)$, an approximate replacement of the longitudinal boost parameter rapidity. The plots are made only for a limited range $(0\leq \eta \leq 4)$, and for the $0-40\%$ most central Au+Au events. We found that at both energies and for each model, $v_2$ as a function of $\eta$ follows more or less similar trend, and the distributions are more or less symmetric. The event-wise averaged $v_2$ peak values consistently exceed the corresponding particle-wise averaged values by $\sim 20\%$. A similar observation has been made by using the AMPT string melting model for $0-40\%$ most central Au+Au collisions at $\sqrt{s_{NN}} = 200$ GeV \cite{Zhou10}. In our case it appears that the systematic differences observed between these two kind of averaging are independent of the models used as well as of the collision energy involved. It is also worth mentioning that when $v_2$ is plotted against $N_{part}$, no significant difference is observed between the two different averaging schemes adopted above.
\subsection{Kinetic radial flow}
\begin{figure}[tbh]
\vspace{-.6cm}
\centering
\includegraphics[height=8cm, width=8.5cm]{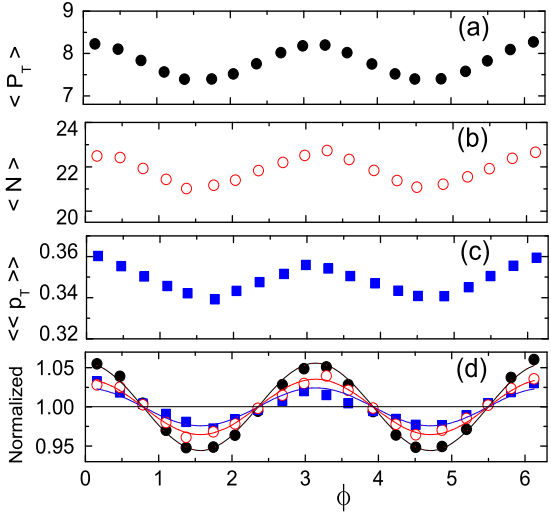}
\vspace{-0.5cm}
\caption{\label{azi_pt} (Color online) Azimuthal distribution of total transverse momentum, mean transverse momentum and multiplicity in Au+Au collisions at $E_{\rm Lab} = 20$A GeV generated by the AMPT (string melting) model.}
\vspace{-0.3cm}
\end{figure}

We now present our results on kinetic radial flow measured in terms of the average transverse momentum of produced charged particles \cite{Ollitrault98}. The azimuthal angle distributions of total transverse momentum $\left<P_{_T}(\phi_m)\right>$ in the $m$-th azimuthal bin
and mean transverse momentum averaged over all particles $\left<\left<p_{_T}(\phi_m)\right>\right>$ in the $m$-th azimuthal bin as introduced respectively, in Eq.(\ref{avg-rap}) and Eq.(\ref{mean-rap}) are obtained for our UrQMD and AMPT (default and string melting) generated Au+Au event samples at $E_{\rm Lab} = 20$A GeV and $40$A GeV. In spite of a significant difference in the collision energy it is found that the distributions are almost similar, and anisotropies are present in the azimuthal distributions of total and mean transverse momentum just like it is also present in the azimuthal distribution of multiplicity. Therefore, $\left<P_{_T}(\phi_m)\right>$ and $\left<\left<p_{_T}(\phi_m)\right>\right>$ distributions only for $E_{\rm Lab} = 20$A GeV are schematically represented in Fig.\ref{azi_pt}(a)--(c) over $0-80\%$ centrality covering the full rapidity range. It is clear that all three distributions exhibit similar periodicity though their amplitudes are different. It should be noted that a similar analysis can be and has already been performed using the transverse rapidity variable \cite{Li2012, Li2013}. It
\begin{figure}[tbh]
\vspace{-.4cm}
\centering
\includegraphics[height=7.2cm,width=9cm]{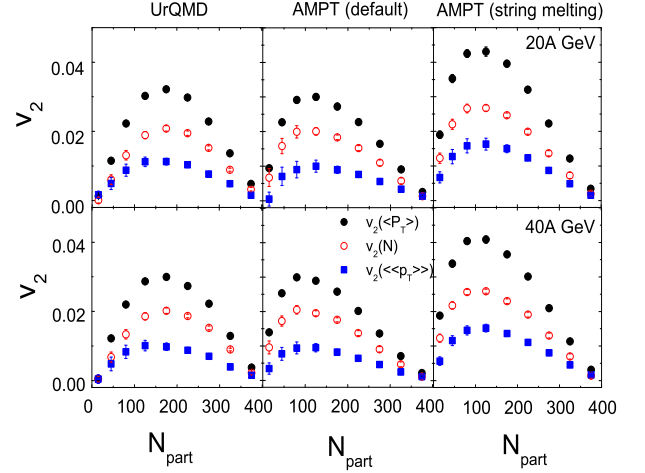}
\vspace{-1cm}
\caption{\label{v2_PT_pT} (Color online) $N_{\rm part}$ dependence of $v_2$ deduced from azimuthal distributions of transverse momentum, multiplicity and mean transverse momentum at 20A GeV and 40A GeV.}
\vspace{-0.1cm}
\end{figure}
is now possible to expand the corresponding distributions in Fourier series as
\begin{eqnarray}
\frac{d\left<P_{_T}\right>}{d\phi} & \approx & v_{_0}\left(\left<P_{_T}\right>\right)\left[1 + 2\,v_{_2}\left(\left<P_{_T}\right>\right)\cos(2\phi)\right]~\mbox{and} \nonumber\\
\frac{d\left<\left<p_{_T}\right>\right>}{d\phi} & \approx & v_{_0}\left(\left<\left<p_{_T}\right>\right>\right)[1 + 2\,v_{_2}\left(\left<\left<p_{_T}\right>\right>\right)\cos(2\phi)]
\label{PT-pT}
\end{eqnarray}
where only the leading terms $(n=0$ and $n=2)$ are retained. It is well known that $v_{_0}$ corresponds to an isotropic part while the anisotropy is quantified by the second Fourier coefficient $v_{_2}$. A combined plot of all three distributions, appropriately normalized and fitted with a dependence like that of Eq.(\ref{PT-pT}) is shown in Fig.\,\ref{v2_PT_pT}(d). As the anisotropy in total transverse momentum $\left<P_{_T}\right>$ contains anisotropies originating both from particle number and mean transverse momentum $\left<p_{_T}\right>$, we observe maximum anisotropy in each $\left<P_{_T}\right>$ distribution. In Fig.\,\ref{v2_PT_pT} we compare the centrality dependence of this elliptic anisotropy obtained from the  distributions of Fig.\,\ref{v2_PT_pT} by plotting $v_2$ against $N_{\rm part}$. At both energies ($E_{\rm Lab}=20$A and $40$A GeV), the centrality dependence are of similar nature, and the extent of anisotropy is consistent with the observation of Fig.\,\ref{v2_PT_pT}(d) i.e., at each centrality $v_2(\left<P_{_T}\right>)$ is larger than $v_{_2}(N_{\rm ch})$, which again is larger than $v_2(\left<\left<p_{_T}\right> \right>)$. Since the gross features of Fig.\,\ref{v2_PT_pT} and Fig.\,\ref{v2-npart} are similar, the elliptic anisotropies are therefore, independent of the underlying distribution i.e., azimuthal angle, multiplicity or transverse momentum, from which they are derived. 
\begin{figure}[tbh]
\vspace{-.4cm}
\includegraphics[height=8cm, width=9cm]{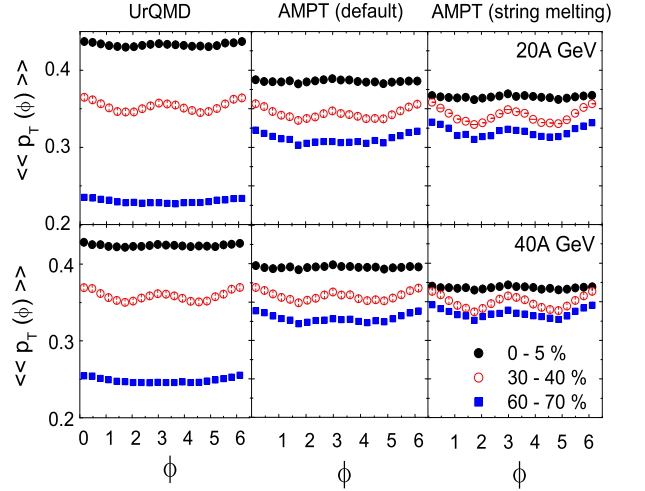}
\vspace{-1.4cm}
\caption{\label{azi_pt_cent} Azimuthal distributions of mean transverse momentum at three different centralities in Au+Au collisions at 20A GeV (top panel) and 40A GeV (bottom panel).}
\end{figure}
In Fig.\,\ref{azi_pt_cent} we show the azimuthal distributions of the mean and total transverse momentum at three different centralities, $0-5\%$, $30-40\%$ and $60-70\%$ for $E_{\rm Lab} = 20$A and $40$A GeV. We find that in mid-central and peripheral collisions the mean transverse momentum distributions are anisotropic, while for central collisions they are approximately isotropic and remain almost uniform with $\phi$ at both energies. In accordance to our expectation, the anisotropy is most prominent in the $30-40\%$ centrality range, be it a multiplicity distribution or transverse momentum distribution. As expected, the radial flow parameter $v_0$ has its highest value in most central collisions of all three event simultors used in this investigation. While in UrQMD the differences in radial flow at different centralities are highest (see the large gaps between different centralities along vertical axis), the corresponding distributions of AMPT (default) are close, while in AMPT (string melting) they are even closer to each other. The features of Fig.\,\ref{azi_pt_cent} are consistent with our observation on centrality dependence of $\left<p_{_T}\right>$ presented in Fig.\,\ref{avgpt}.
\section{Summary}
\label{Summary}
We have performed a simulation study and presented the results obtained thereof in this paper on azimuthal anisotropy, particularly on the elliptic flow of charged particles produced in symmetric Au+Au system at $E_{\rm Lab}=20$A and $40$A GeV incident energies using event generators like UrQMD, AMPT(default) and AMPT (string melting). The collision energies are so chosen as to match with the energy values to be available in the upcoming CBM experiment at FAIR. The simulated results presented in this paper will allow us to pin point our expectations regarding the collective behavior of particles coming out of baryon rich fireballs produced in heavy-ion interactions at FAIR energies, and compare them with the experiments where similar as well as at significantly different collision energies are involved. Major observations of the present analysis are summarized below.\\
\\
Anisotropies in the azimuthal angle distributions of charged particles are observed in Au+Au events at $E_{\rm Lab}=20$A and $40$A GeV generated by the UrQMD, the AMPT(default) and the AMPT (string melting) models. The anisotropy is maximum in the mid-central collisions and in events generated by the AMPT string melting model. Dependence of elliptic flow parameter on $N_{\rm part}$, $p_{_T}$ and $\eta$ of the emitted charged particles do not show any unusual behavior. If the effect of nuclear geometry (overlapping part of the colliding nuclei) is taken care of, the elliptic flow parameter rises monotonically with particle density in the transverse plane. Initial spatial anisotropy also results in kinetic radial expansions that are quantified by the $v_2$ values obtained from $\left<P_{T}\right>$ and $\left<\left<p_{_T}\right>\right>$ distributions. We see that the elliptic flow is maximum in the $\left<P_{T}\right>$ distribution, where contribution from both particle multiplicity and $\left<\left<p_{_T}\right>\right>$ distributions are present. While the kinetic radial flow itself is always maximum at highest centralities, its elliptic anisotropy is largest in mid-central collisions of the AMPT string melting model.\\
\\
The collective behavior of particles belonging to different species coming out of symmetric nucleus-nucleus collisions in the $E_{\rm Lab}=10-40$A GeV energy needs to be critically examined. As far as the analysis part is concerned, a lot of progress has taken place during the last several years. In the absence of experiments they can be easily implemented to simulated data. In view of the upcoming CBM program this is extremely significant and will serve as a precursor to the real experimental results.\\
\section{Acknowledgement} One of us (AM) would like to acknowledge the financial support provided by the University of North Bengal during the period of this investigation. 

\end{document}